\documentstyle[twocolumn,prc,aps]{revtex}
\input epsf
\begin{document}
\twocolumn[\hsize\textwidth\columnwidth\hsize\csname @twocolumnfalse\endcsname 


\title{Kaon and Pion Fluctuations from Small Disoriented Chiral Condensates}
\author{Sean Gavin$^a$ and Joseph I. Kapusta$^b$} 
\address{
a) Department of Physics and Astronomy, Wayne State University, 
Detroit, MI, 48202\\
b) School of Physics and Astronomy, University of Minnesota,
Minneapolis, MN 55455}


\date{\today}
\maketitle
\begin{abstract}
Enhancement of $\Omega$ and $\overline\Omega$ baryon production in
Pb+Pb collisions at a c.m.\ energy of 17 $A$ GeV can be explained by
the formation of many small disoriented chiral condensate
regions. This explanation implies that neutral and charged kaons as
well as pions must exhibit novel isospin fluctuations. We compute the
distribution of the fraction of neutral pions and kaons from such
regions. We then propose robust statistical observables that can be
used to extract the novel fluctuations from background contributions
in $\pi^0\pi^\pm$ and $K_S^0K^\pm$ measurements at RHIC and LHC.

\vspace{0.1in}
\end{abstract}
]

\begin{narrowtext}

\newpage

\section{Introduction}

Heavy ion collisions at the Brookhaven Relativistic Heavy Ion Collider (RHIC) 
at center of mass energies up to $200$~$A$~GeV and the CERN Large Hadron 
Collider (LHC) at $5.5$~$A$~TeV may produce matter in which chiral symmetry is 
restored. One possible consequence of the restoration and the subsequent 
re-breaking of chiral symmetry is the formation of disoriented chiral condensates 
(DCC) -- transient regions in which the average chiral order parameter differs 
from its value in the surrounding vacuum \cite{DCCreview,qmrev1,qmrev2}. 

Measurements of $\Omega$ and $\overline\Omega$ baryon enhancement 
\cite{SPSbaryons} at $17$~$A$~GeV at the CERN SPS can be explained by the 
production of many small DCC regions within individual collision events 
\cite{KapustaWong}. If true, this explanation has two important consequences. 
First, the DCC regions must be rather small, with a size of about $2$~fm. Such a 
size is consistent with predictions based on dynamical simulations of the two 
flavor linear sigma model \cite{GGP1}. More startling is the second implication 
that the evolution of the condensate can have a significant effect on {\it 
strange} particle production. The importance of strange degrees of freedom in 
describing chiral restoration has been long appreciated 
\cite{PisarskiWilczek,Columbia,GGP2,Karsch,Lenaghan}, but simulations of the 
three flavor linear sigma model had suggested that strange kaon fields
are much less important than the pion fields~\cite{Randrup}. Nevertheless, 
the $\Omega$ and $\overline\Omega$ data demand that we explore without prejudice 
techniques for measuring kaon fluctuations.

In this paper we study pion and kaon isospin fluctuations in the
presence of many small DCC. In the next section we compute probability
distributions that describe the DCC contribution to these
fluctuations. Pion fluctuations due to many small DCC have been
addressed by Amado and Lu \cite{AmadoLu} and Chow and Cohen
\cite{ChowCohen}, although the distribution we compute is new. Ours
is the first work to study kaon fluctuations. In sec.~3 we combine
the DCC fluctuations with a contribution from a random thermal
background. In sec.~4 we discuss how the size and number of DCC vary
with impact parameter, target and projectile size. In sec.~5 we assess
robust statistical observables that can be used to measure the impact
of many small DCC at RHIC and LHC. In particular, we obtain a dynamic
isospin fluctuation observable analogous to the dynamic charge
observable used to measure net charge fluctuations at RHIC \cite{Voloshin}. 
Of the quantities considered, this observable isolates the DCC effect from
other sources of fluctuations best.

To illustrate how a strange DCC can form, we first consider QCD with
only up and down quark flavors. Equilibrium high temperature QCD
respects chiral symmetry if the quarks are taken to be massless. This
symmetry is broken below $T_c\sim 150$~MeV by the formation of a
chiral condensate $\langle \sigma\rangle \sim \langle
\overline{u}u+\overline{d}d\rangle$ that is a scalar isopin singlet. 
However, chiral symmetry implies that $\sigma$ is degenerate with a
pseudoscalar isospin triplet of fields with the same quantum numbers
as the pions. In reality, chiral symmetry is only approximate and the
140~MeV pion mass is different from the $800\pm 400$~MeV mass of the
leading sigma candidate \cite{pdg}. Nevertheless, lattice calculations
exhibit a dramatic drop of $\langle \sigma\rangle$ near $T_c$ at
finite quark masses.

A DCC can form when a heavy ion collision produces a high energy
density quark-gluon system that then rapidly expands and cools through
the critical temperature. Such a system can initially break chiral
symmetry along one of the pion directions, but must then evolve to the
$T=0$ vacuum by radiating pions. A single coherent DCC radiates a
fraction $f_\pi$ of neutral pions compared to the total that satisfies
the probability distribution
\begin{equation}
\rho_1(f_\pi) = {{1}\over{2f_\pi^{1/2}}}\,\,\,\,\,\,\,\,\,\,\,\,\,\,\,\, 0 < 
f_\pi \le 1,
\end{equation}
\cite{Anselm,Blaizot,Bjorken}. Such isospin fluctuations constitute the primary 
signal for DCC formation. The enhancement of baryon-antibaryon pair
production is a secondary effect due to the relation between baryon
number and the topology of the pion condensate field \cite{KapustaSrivastava}.

This two flavor idealization only applies if the strange quark mass
$m_s$ can be taken to be infinite. Alternatively, if we take $m_s =
m_u = m_d =0$, then the chiral condensate would be an up-down-strange
symmetric scalar field. The more realistic case of $m_s\sim 100$~MeV
is between these extremes, so that $\langle \sigma\rangle \sim \langle
\cos\theta(\overline{u}u+\overline{d}d) +
\sin\theta(\overline{s}s)\rangle$. The mixing angle $\theta$ is highly
uncertain since it depends on the sigma mass together with the $\pi,
K, \eta$ and $\eta^\prime$ masses and the $\eta-
\eta^\prime$ mixing angle \cite{GGP2}. A disoriented condensate can evolve by 
radiating $\pi, K, \eta$ and $\eta^\prime$ mesons, with the neutral pion 
fraction satisfying (1). Randrup and Sch\"affner-Bielich find that the kaon 
fluctuations from a single large DCC satisfy \cite{Randrup}
\begin{equation}
\rho_1(f_K) = 1\,\,\,\,\,\,\,\,\,\,\,\,\,\,\,\, 0 \le f_K \le 1,
\end{equation}
where $f_K = (K^0 + \overline{K}^0)/(K^+ + K^- + K^0 +
\overline{K}^0)$. Moreover, the condensate fluctuations can now produce
strange baryon pairs \cite{KapustaWong}. Linear sigma model simulations indicate that
pion fluctuations dominate three-flavor DCC behavior, while the
fraction of energy imparted to kaon fluctuations is very small due to
the kaons' larger mass. On the other hand, domain formation may be induced by other
mechanisms such as bubble formation \cite{qm95} or decay of the Polyakov loop condensate
\cite{Pisarski}.  

Why does the DCC's size matter? Pion measurements in individual
collision events can distinguish DCC isospin fluctuations from a
thermal background only if the disoriented region is sufficiently
large \cite{qmrev1}. DCC can then be the dominant source of pions at
low transverse momenta, since $\langle p_t \rangle\sim 1/R$ for a
coherent region of size $R$. Experiments focusing on low $p_t$ can
study neutral and charged pion fluctuations \cite{Bjorken}, wavelet
\cite{Ina} and HBT signals \cite{qmrev1,HBT} to extract detailed information. In 
contrast, for small domains ($R<3$~fm \cite{qmrev1}) DCC signals are
hidden by fluctuations due to ordinary incoherent production
mechanisms. This holds even if many such regions are produced per
event. DCC mesons from small regions may have momenta of a few hundred
MeV, nearer the $pp$ mean value. Different regions would not add
coherently to alter HBT, nor would their small spatial structures
affect wavelet analyses.

Importantly, baryon pair enhancement is substantial only if there are
many small incoherent regions. The large winding numbers that produce
baryon-antibaryon pairs require many small regions with random
relative orientations of the pion field. To describe strange
antibaryon enhancement, Kapusta and Wong assume roughly 100 DCC
regions of size of roughly $2$~fm \cite{KapustaWong}. Topological models
of baryon-antibaryon pair production successfully describe $e+e-$ and
hadronic collision data \cite{EllisKowalski}. The connection of DCC to
topological pair production was pointed out in
Ref.~\cite{KapustaSrivastava}; see also \cite{DeGrand}.

\section{Fluctuations in Neutral DCC Mesons}

In this section we will compute the statistical distribution of the
ratio of neutral to total number of mesons, first for kaons and then
for pions. In both cases the limit that the number of DCC domains
becomes large is taken. It is natural that this limit results in a
Gaussian distribution for both kaons and pions on account of the
Central Limit Theorem.  In the next section these distributions will
be folded together with a random or thermal source which most likely
would comprise the bulk of the mesons in a high energy heavy ion
collision.

\subsection{Kaons}

Define $f = (K^0 + \overline{K}^0)/(K^+ + K^- + K^0 + \overline{K}^0)$.  To an 
excellent approximation the number of neutral kaons is equal to twice the number 
of short-lived neutral kaons $K_S$ which are more readily measurable in high 
energy heavy ion collisions.  The fraction $f$ ranges from 0 to 1.

The statistical distribution in $f$ for a single domain is $\rho_1(f)
= 1$.  The distribution for $n$ randomly oriented, independent domains
is
\begin{equation}
\rho_n(f) = \int  \prod_{k=1}^n df_k \, 
\rho_1(f_k) \, \delta\left( f - \frac{1}{n} \sum_{j=1}^n f_j \right) \, .
\end{equation}
The Dirac delta function can be represented as an integral.  Then $\rho_n$ can 
be written as
\begin{equation}
\rho_n(f) = \int_{-\infty}^{\infty} \frac{dz}{2\pi} e^{-i fz}
\left[ \int_0^1 dx \rho_1(x) e^{ixz/n} \right]^n \, .
\end{equation}
Since $\rho_1(f) = 1$, the integration over $x$ can be done, resulting in a one-
dimensional integral.
\begin{equation}
\rho_n(f) = \frac{n}{\pi} \int_{-\infty}^{\infty} dt \left( \frac{\sin t}{t}
\right)^n \cos\left[n(1-2f)t\right]
\end{equation}
The integral can be evaluated and expressed in terms of a finite sum.
\begin{equation}
\rho_n(f) = n^2 \sum_{0 \leq k < n(1-f)} (-1)^k
\frac{[n(1-f)-k]^{n-1}}{k! (n-k)!}
\end{equation}

It is useful to have a simple analytic formula for $\rho_n$ in the limit
that $n \gg 1$.  In this limit the factor $(\sin t/t)^n$ in the integral 
formula is strongly peaked at $t=0$.  Let us write this factor as
$\exp(-F(t))$ with a view towards a saddle point approximation.  We get
\begin{eqnarray}
F(0) &=& F'(0) = 0 \nonumber \\
F''(0) &=& n/3 \nonumber \\
e^{-F(t)} &\approx & e^{-nt^2/6} \, .
\end{eqnarray}
Use of this approximation yields the asymptotic formula
\begin{equation}
\rho_n(f) = \sqrt{\frac{6n}{\pi}} \exp [ -6n(f-1/2)^2] \, .
\end{equation}
The distribution is strongly peaked around $f=1/2$ as one might expect.

\begin{figure} 
\epsfxsize=3.0in
\centerline{\epsffile{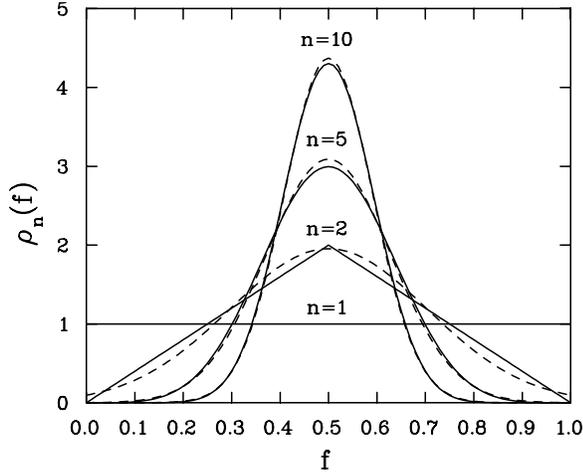}}
\vskip +0.1in
\caption[]{The probability distribution for the ratio
of neutral to total number of DCC kaons from $n$ domains.
The dashed curves represent Gaussian distributions.  From eqs. (6)
and (8).
  }\end{figure}
Figure 1 shows the evolution of $\rho_n(f)$ with $n$.  It goes from a flat 
distribution for $n=1$ to a Gaussian sharply peaked at $f=1/2$ as $n$ becomes 
large compared to 1.  In fact a Gaussian is a very good representation for
$n > 2$.

\subsection{Pions}

Define $f = \pi^0/(\pi^+ + \pi^- + \pi^0)$.  To a good approximation the number 
of neutral pions is equal to half the number of photons.   Therefore, to this 
level of precision, it is not necessary to identify each $\pi^0$ via its decay 
into $2\gamma$.  The fraction $f$ ranges from 0 to 1.

The statistical distribution in $f$ for a single domain is
$\rho_1(f) = 1/2\sqrt{f}$.
The distribution for $n$ randomly oriented, 
independent domains can be computed along the same lines as for kaons.
\begin{equation}
\rho_n(f) = \int_{-\infty}^{\infty} \frac{dz}{2\pi} e^{-i fz}
\left[ \int_0^1 \frac{dx}{2\sqrt{x}} e^{ixz/n} \right]^n \, .
\end{equation}
Since pions from DCC have been extensively studied we shall be content to 
evaluate the distribution in the large $n$ limit.  This is accomplished by 
expanding the exponential in the $x$ integration to second order in $1/n$, 
evaluating the resulting integrals, and exponentiating.  Thus
\begin{equation}
\int_0^1 \frac{dx}{2\sqrt{x}} e^{ixz/n} \approx \exp\left[
i\frac{z}{3n} - \frac{2z^2}{45n^2} \right] \, .
\end{equation}
The $z$ integral can then be done, yielding a Gaussian centered at $f=1/3$.
\begin{equation}
\rho_n(f) = \sqrt{\frac{45n}{4\pi}} \exp \left[ -45n(f-1/3)^2/4 \right]
\end{equation}

\section{Folding DCC and Thermal Mesons}

In a more realistic scenario some kaons will come from the decay or realignment 
of DCC domains and some will come from more conventional sources.  We shall 
refer to the latter as random or thermal, even though that may be a bit of a 
misnomer.  What we mean by random or thermal is that the distribution of kaons 
from non-DCC sources is
\begin{equation}
\rho_0(f_0) = \frac{1}{2\pi \sigma_0^2} \exp \left[ -(f_0-1/2)^2
/2\sigma_0^2 \right] \, .
\end{equation}
For a completely random source the width $\sigma_0$ is related to the total 
number $N_{\rm random}$ of non-DCC kaons by
\begin{equation}
\sigma^2_0 = \frac{1/2(1-1/2)}{N_{\rm random}} = \frac{1}{4N_{\rm random}} \, .
\end{equation}

Now let us assume that a fraction $\alpha_K$ of all kaons come from non-DCC 
sources and the remaining fraction $\beta_K = 1-\alpha_K$ come from $n \gg 1$ 
independent DCC domains.  Letting $N$ denote the total number of 
kaons, we have $N_{\rm random} = \alpha_K N$ and $N_{\rm DCC} = \beta_K N$.  
Folding together two Gaussians gives a Gaussian.
\begin{eqnarray}
\rho_K(f) &=& \int df_0 df_n \rho_0(f_0) \rho_n(f_n) \delta (f - \alpha_K f_0
-\beta_K f_n) \nonumber \\
&=& \frac{1}{\sqrt{2\pi \Delta_K^2}} \exp \left[ -(f-1/2)^2/2\Delta_K^2 \right]
\end{eqnarray}
The net width is
\begin{equation}
\Delta_K^2 = \frac{\alpha_K}{4N} + \frac{\beta_K^2}{12n} =
\frac{1}{4N} + \left\{ \frac{\beta_K^2}{12n} - \frac{\beta_K}{4N}
\right\}\, .
\end{equation}
The expression in curly brackets at the end represents the difference between 
the actual width and the width the distribution would have if there was no 
contribution from DCC kaons.  This change in the width may be positive or 
negative, depending on the parameters.

An analogous analysis can be given for pions.  This results in the distribution
\begin{eqnarray}
\rho_{\pi}(f) &=& \int df_0 df_n \rho_0(f_0) \rho_n(f_n) \delta (f - 
\alpha_{\pi} f_0 -\beta_{\pi} f_n) \nonumber \\
&=& \frac{1}{\sqrt{2\pi \Delta_{\pi}^2}} 
\exp \left[ -(f-1/3)^2/2\Delta_{\pi}^2 \right]
\end{eqnarray}
with a net width of
\begin{equation}
\Delta_{\pi}^2 = \frac{2\alpha_{\pi}}{9N} + 
\frac{2\beta_{\pi}^2}{45n} =
\frac{2}{9N} + \left\{ \frac{2\beta_{\pi}^2}{45n} - 
\frac{2\beta_{\pi}}{9N} \right\}\, .
\end{equation}
As with the kaons, the last expression in curly brackets represents the 
difference between the actual width and the width the distribution would have if 
there was no contribution from DCC pions.  Note that the fractions $\alpha_K$ 
and $\alpha_{\pi}$ need not be the same.

\section{Volume or Surface Scaling?}

The issue we wish to address is whether the number of DCC mesons (kaons 
or pions) scales with the volume or surface area of the system.  This is an 
important issue when studying the impact parameter dependence or the dependence 
on the size of the projectile and target nuclei.

In this paper it is assumed that DCC have a typical size of order 2 fm which 
does not change much with collision energy or the total volume of the system.  
Thus the number of domains $n$ is just given by the ratio of the two volumes: $n 
= V_{\rm system}/V_{DCC}$.  Scaling of the number of DCC kaons or pions, 
$N_{DCC}$, with $n$ or $V_{\rm system}$ is the same because the size of 
individual domains is fixed.  The $N_{DCC}$ will depend on the extra energy 
associated with the formation of a domain.  If the up and down quark masses were 
zero then QCD would have perfect SU(2) flavor symmetry.  In that case the energy 
density of a large uniform domain would be independent of its orientation.  All 
directions in chiral space are equivalent.  However, the misalignment between 
adjacent domains results in a surface energy, and so the number of DCC pions 
would be proportional to the total surface energy between domains.  The up and 
down quark masses are not zero (they're about 5 to 7 MeV), and this will result 
in an excess volume energy too.   For pions one might expect the surface energy 
to dominate since the up and down quark masses are so small.  For kaons one 
might expect the volume and surface energy contributions to be comparable on 
account of relatively large mass of the strange quark (about 120 MeV).

Let us analyze the scaling of the excess surface energy more quantitatively.
Consider a cube with sides of length $L$ into which fit $n = (L/l)^3$ cubic 
domains, each with sides of length $l$.  Assuming that each domain is oriented 
independently of its neighbors, the total surface energy scales with the total 
surface area, which is $3\left(\frac{L}{l}+1\right) \frac{L^2}{l^2}$.  With $L 
\gg l$, and with the domain size $l$ fixed, the total surface area, energy, and 
therefore number of DCC mesons scale with $n$ to the power of one.  Thus 
$N_{DCC} \propto n \propto V_{\rm system}$ no matter whether one imagines the 
excess energy being associated with domain interfaces or with domain interiors.

\section{Statistical Analysis}

Detection of small incoherent DCC regions in high energy heavy ion collisions 
requires a statistical analysis in the $\pi^0\pi^\pm$ or the $K_S^0K^\pm$ 
channels. Neutral mesons can be detected by the decays $\pi^0\rightarrow 
\gamma\gamma$ or $K^0_S\rightarrow \pi^+\pi^-$. The analysis we propose is 
sensitive to correlations due to isospin fluctuations.  We expect these 
correlations to vary when DCC regions increase in abundance or size as 
centrality, ion-mass number $A$, or beam energy are changed. Correlation results 
combined with other signals, such as baryon enhancement \cite{KapustaWong}, can 
be used to build a circumstantial case for DCC production. 

Correlations of $\pi^0\pi^\pm$ and $ K_S^0 K^\pm$ can be determined by
measuring the robust isospin covariance,
\begin{equation}
R_{c0} = {{\langle N_cN_0\rangle - \langle N_c\rangle\langle N_0\rangle}
\over{\langle N_c\rangle\langle N_0\rangle}},
\end{equation}
where $N_0$ and $N_c$ are the number of neutral and charged mesons. We
take $N_0= N_{\pi^0}$ and $N_c=N_{\pi^+}+N_{\pi^-}$ for pion
fluctuations and $N_0=2N_{ K^0_S}$ and $N_c=N_{K^+}+N_{K^-}$ for kaon
fluctuations. The ratio (18) has two features that are convenient for
experimental determination. First, this observable is independent of
detection efficiency as are the ``robust'' ratios discussed in
\cite{Taylor}. Robust observables are useful for DCC studies because
charged and neutral particles are identified using very different
techniques and, consequently, are detected with different
efficiency. Observe that robust quantities are not affected by the
unobserved $K_L^0$, since the strong-interaction eigenstates $K^0$ and
${\overline K}^0$ are a superposition $K_L^0$ and $K_S^0$ until their
decay well outside the collision region.  Second, since (18) is
obtained from a statistical analysis, individual $\pi^0\rightarrow
\gamma\gamma$ or $K^0_S\rightarrow \pi^+\pi^-$ need not be fully
reconstructed in each event. This feature is crucial because it would
be extraordinarily difficult -- if not impossible -- to reconstruct a
low momentum $\pi^0$ in heavy ion collisions except on a statistical
basis.

Next we define robust variance
\begin{equation}
R_{aa} = {{\langle N_a^2\rangle - \langle N_a\rangle^2 - \langle N_a\rangle}
\over{\langle N_a\rangle^2}},
\end{equation}
where $a = c$ or 0. To see why (19) is robust, denote the
probability of detecting each meson $\epsilon$ and the probability
of missing it $1-\epsilon$. For a binomial distribution the average number of 
measured particles is $\langle N_a\rangle^{\rm exp} = \epsilon\langle 
N_a\rangle$ while the average square is $\langle N_a^2\rangle^{\rm exp} = 
\epsilon^2\langle
N_a^2\rangle+\epsilon(1-\epsilon)\langle N_a\rangle$. We then find
\begin{equation}
R_{aa}^{\rm exp} = R_{aa},
\end{equation}
independent of $\epsilon$ \cite{Pruneau}; the proof that (18) is
robust is similar. The ratios (18) and (19) are strictly robust only
if the efficiency $\epsilon$ is independent of multiplicity. Further
properties and advantages of these and similar quantities are
discussed in \cite{Pruneau}.

To study DCC fluctuations we define the dynamic isospin observable
\begin{equation}
\nu_{\rm dyn}^{c0} = R_{cc} + R_{00} - 2R_{c0}.
\end{equation}
Analogous observables have been employed to study net charge
fluctuations in particle physics \cite{Whitmore,Boggild} and were
considered in a heavy ion context in \cite{Voloshin} and
\cite{Mrowczynski}. This quantity can be written in terms of
\begin{equation}
\nu^{c0} = \left\langle\left({{N_0}\over{\langle N_0\rangle}} -
{{N_c}\over{\langle N_c\rangle}} \right)^2\right\rangle.
\end{equation}  
To isolate the dynamical isospin fluctuations from other sources
of fluctuations, one obtains (21) by subtracting from (22) the
uncorrelated Poisson limit $\nu_{\rm stat}^{c0}=\langle
N_0\rangle^{-1}+\langle N_c\rangle^{-1}$.  Indeed, we show in (27)
below that the quantity (21) depends primarily on the fluctuations of
the neutral fraction $f$, while the individual ratios (18) and (19)
have additional contributions.

We illustrate the effect of DCC on the dynamic isospin fluctuations by
writing $N_0 = fN$ and $N_c = (1-f)N$.  Small fluctuations on $f$ or $N$
results in the changes
\begin{eqnarray}
\frac{\Delta N_0}{\langle N_0 \rangle} &=& {{\Delta N}\over{\langle N\rangle}}
+ {{\Delta f}\over{\langle f\rangle}} \, ,\nonumber \\ 
\frac{\Delta N_c}{\langle N_c \rangle} &=& {{\Delta N}\over{\langle N\rangle}}
 -  {{\Delta f}\over{1-\langle 
f\rangle}} \, . 
\end{eqnarray}
We obtain the average
\begin{equation}
{{\langle\Delta N_0^2\rangle}\over{\langle N_0\rangle^2}}
 = v + {{2c}\over{\langle N\rangle \langle
f\rangle}} + {{\Delta^2}\over{\langle f\rangle^2}} \, .
\end{equation}
Here the contribution of the variance of the total number of mesons
is $v \equiv \langle \Delta N^2\rangle/\langle N\rangle^2$ and the
charge-total covariance is $c \equiv \langle \Delta N\Delta f\rangle$.
DCC formation primarily effects the charge fluctuation contribution,
$\Delta^2 \equiv \langle (\Delta f)^2\rangle$, from (15) or (17). Similarly,
\begin{equation}
{{\langle\Delta N_c^2\rangle}\over{\langle N_c\rangle^2}} 
= v - {{2c}\over{\langle N\rangle(1-\langle
f\rangle)}} + {{\Delta^2}\over{(1-\langle f\rangle)^2}} \, ,
\end{equation}
and
\begin{equation}
R_{c0}= v + \left({{1}\over{\langle f\rangle}} - {{1}\over{1-\langle 
f\rangle}}\right){{c}\over{\langle N\rangle}}
- {{\Delta^2}\over{(1-\langle f\rangle)^2}} \, 
\end{equation}
where $R_{c0}$ is given by (18). Using (21) we get 
\begin{equation}
\nu_{\rm dyn}^{c0} = {{1}\over{\langle f\rangle(1-\langle f\rangle)}}
\left({{\Delta^2}\over{\langle f\rangle(1-\langle 
f\rangle})} -{{1}\over{\langle N\rangle}}\right).
\end{equation}
This observable isolates the isospin fluctuations, whereas the
individual $R_{ab}$ depend on the fluctuations in total meson number,
$v$ and $c$ as well.

We estimate the effect of DCC on the dynamical fluctuations (27) using
(15) and (17). We take $\langle N\rangle = N_K$ for kaons and $\langle N\rangle 
= N_{\pi}$ for pions; these are the total number of mesons of the indicated 
kind.  For kaons
\begin{equation}
\nu_{\rm dyn}^{c0}({\rm K~DCC}) = 4\beta_K \left( \frac{\beta_K}{3n}
- \frac{1}{N_K} \right) \, ,
\end{equation}
and for pions
\begin{equation} 
\nu_{\rm dyn}^{c0}(\pi~{\rm DCC}) = 4.5 \beta_{\pi} \left(
\frac{\beta_{\pi}}{5n} - \frac{1}{N_{\pi}} \right) \, .
\end{equation}
These quantities can be positive or negative depending on the
magnitude of $\beta$ compared to the number of domains per kaon. In
fact the dynamical fluctuation may even be positive for one kind of
meson and negative for the other. 

How big is the DCC effect compared to alternative sources of
fluctuations? In the absence of DCC $\alpha = 1$ and $\beta = 0$ so
that (29) implies $\nu_{\rm dyn}^{c0} \equiv 0$ for both pions
and kaons. On the other hand, in models which treat nuclear collisions
as a superposition of independent nucleon-nucleon collisions, each
nucleon-nucleon collision contributes an amount $\nu_{c0}^{pp}$ to the overall
fluctuations. Consequently, $M$ nucleon-nucleon collisions can contribute an 
amount $\nu_{c0}^{pp}/M$ to the total $\nu_{\rm dyn}^{c0}$
\cite{hamlet}.  While little is known from $pp$ experiments about kaon
fluctuations, HIJING and RQMD models yield negative values of
$\nu_{c0}^{pp}$. For kaons, HIJING simulations of central Au+Au at 200
$A$ GeV yield $\nu_{\rm dyn}^{c0} \approx -0.002$ for 47 $K^+$ and 44
$K_S^0$ on average \cite{HIJING}.  The onset of a DCC contribution to
$\nu_{\rm dyn}^{c0}$ can substantially change this value. A detailed
of analysis of this problem within microscopic models will appear
elsewhere \cite{hamlet}.

\section{Discussion and Conclusion}

Reference \cite{KapustaWong} argued that the anomalous abundance and
transverse momentum distributions of $\Omega$ and $\overline{\Omega}$
baryons in central collisions between Pb nuclei at 17 $A$ GeV at the
CERN SPS is evidence that they are produced as topological defects
arising from the formation of many domains of disoriented chiral
condensates (DCC) with an average domain size of about 2 fm. Motivated
by this interpretation, we have studied the effect of DCC on the
distribution of the fractions of neutral kaons and pions.  We showed that
the distributions are accurately described by Gaussians with centroids
at $f$ = 1/2 and 1/3, respectively, once the number of domains exceeds
just a few.  Folding together kaons or pions arising from DCC with
other sources that are Gaussian distributed results once again in
Gaussians. These may have a width that is greater or less than a
purely random source without DCC formation.

The DCC pioneers \cite{Anselm,Blaizot,Bjorken,DCCreview} had hoped
that a large percentage of pions might be emitted from just a few big
domains, on the order of 5 to 8 fm (kaons were not considered). Such
large domains have been ruled out at SPS \cite{qmrev2}, but remain
possible at RHIC. More conservatively, as the number of domains grow
and their average size diminishes, the impression left on the
fluctuations in the neutral fraction becomes more subtle and less
unique.  For many small domains, statistical measurements of both
neutral kaons (pions) and charged kaons (pions) are needed.  Since not
every hadron emitted can possibly be detected with 100\% efficiency,
and since the experimental techniques that identify $K_S$, $K^{\pm}$,
$\pi^0$ and $\pi^{\pm}$ are very different, we have identified robust
observables that are essentially independent of all these
uncertainties.  In particular, we propose that the dynamical isospin
observable (21) can be parameterized as in eqs. (28) and (29). DCC
effects can appear as changes in the magnitude of the dynamical
isospin observable as centrality is varied. We emphasize that similar
consequence may follow from any mechanism that produces many small
domains that decay to pions and kaons, such as the Polyakov Loop
Condensate \cite{Pisarski}. We anxiously await what RHIC will have to
say!

\section*{Acknowledgements}

S.~G.~thanks the Nuclear Theory Group at the University of Minnesota
for kind hospitality during visits in March and December 2001 during
which much of this work was done.  This work was supported by the
U.S. Department of Energy under grant numbers DE-FG02-87ER40328 and
DE-FG02-92ER40713.

\end{narrowtext}

\begin{thebibliography}{99}
\bibliographystyle{unsrt}

\bibitem{DCCreview} K.~Rajagopal and F.~Wilczek, Nucl.\ Phys.\
{\bf B399}, 395 (1993);
{\bf B404}, 577 (1993); 
K.~Rajagopal, in Quark-Gluon Plasma 2, R.~C.~Hwa ed., (World Scientific, 1995); hep-ph/9504310.

\bibitem{qmrev1} S.~Gavin, Nucl.\ Phys.\ {\bf A590}, 163c (1995).

\bibitem{qmrev2} T.~Nayak, {\it et al.} (WA98 Collaboration), 
Nucl.\ Phys.\ {\bf A638}, 249c (1998).

\bibitem{SPSbaryons}
J.~B.~Kinson,
J.\ Phys.\ G {\bf 25}, 143 (1999).

\bibitem{KapustaWong} 
J.~I.~Kapusta and S.~M.~Wong,
Phys.\ Rev.\ Lett.\  {\bf 86}, 4251 (2001); nucl-th/0012006.

\bibitem{GGP1}
S.~Gavin, A.~Gocksch and R.~D.~Pisarski,
Phys.\ Rev.\ Lett.\  {\bf 72}, 2143 (1994); hep-ph/9310228.

\bibitem{PisarskiWilczek}
R.~D.~Pisarski and F.~Wilczek,
Phys.\ Rev.\ D {\bf 29}, 338 (1984).

\bibitem{Columbia}
F.~R.~Brown {\it et al.},
Phys.\ Rev.\ Lett.\  {\bf 65}, 2491 (1990).

\bibitem{GGP2}
S.~Gavin, A.~Gocksch and R.~D.~Pisarski, Phys.\ Rev.\ D {\bf 49}, 3079 (1994); hep-ph/9311350.

\bibitem{Karsch}
C.~Schmidt, F.~Karsch and E.~Laermann, hep-lat/0110039.

\bibitem{Lenaghan}
J.~T.~Lenaghan, D.~H.~Rischke and J.~Schaffner-Bielich,
Phys.\ Rev.\ D {\bf 62}, 085008 (2000); nucl-th/0004006.

\bibitem{Randrup} J.~Schaffner-Bielich and J.~Randrup, Phys.\ Rev.\ C
{\bf 59}, 3329 (1999); nucl-th/9812032.

\bibitem{AmadoLu}
R.~D.~Amado and Y.~Lu,
Phys.\ Rev.\ D {\bf 54}, 7075 (1996); hep-ph/9608242.

\bibitem{ChowCohen}
C.~K.~Chow and T.~D.~Cohen,
Phys.\ Rev.\ C {\bf 60}, 054902 (1999); nucl-th/9908013.

\bibitem{Voloshin}
S.~A.~Voloshin  [STAR Collaboration], nucl-ex/0109006.

\bibitem{pdg}
D.~E.~Groom {\it et al.}  [Particle Data Group Collaboration],
Eur.\ Phys.\ J.\ C {\bf 15}, 1 (2000).

\bibitem{Anselm} A.~A.~Anselm and M.~G.~Ryskin, Phys.\ Lett.\ {\bf B266},
482 (1991).

\bibitem{Blaizot} J.-P.~Blaizot and A.~Kryzywicki, Phys.\ Rev.\ D
{\bf 46}, 246 (1992).

\bibitem{Bjorken} J.~D.~Bjorken, K.~L.~Kowalski and C.~C.~Taylor,
Report SLAC-PUB-6109 (1993), hep-ph/9309235, unpublished.

\bibitem{KapustaSrivastava}
J.~I.~Kapusta and A.~M.~Srivastava,
Phys.\ Rev.\ D {\bf 52}, 2977 (1995); hep-ph/9404356.

\bibitem{qm95} J.~I.~Kapusta, A.~P.~Vischer and R.~Venugopalan,
Phys.\ Rev.\ C {\bf 51}, 901 (1995); nucl-th/9408029; J.~I.~Kapusta and  A.~P.~Vischer,
Z.\ Phys.\ {\bf C75}, 507 (1997); nucl-th/9605023.

\bibitem{Pisarski}
A.~Dumitru and R.~D.~Pisarski,
Phys.\ Lett.\ {\bf B504}, 282 (2001); hep-ph/0010083.

\bibitem{Ina}
Z.~Huang, I.~Sarcevic, R.~Thews and X.~N.~Wang,
Phys.\ Rev.\ D {\bf 54}, 750 (1996); hep-ph/9511387.

\bibitem{HBT}
H.~Hiro-Oka and H.~Minakata,
Phys.\ Lett.\ {\bf B425}, 129 (1998)
[Erratum-ibid.\ {\bf B434}, 461 (1998)]; hep-ph/9712476.

\bibitem{EllisKowalski}
J.~R.~Ellis and H.~Kowalski,
Phys.\ Lett.\ {\bf B214}, 161 (1988);
Nucl.\ Phys.\ {\bf B327}, 32 (1989).

\bibitem{DeGrand}
T.~A.~DeGrand,
Phys.\ Rev.\ D {\bf 30}, 2001 (1984);
J.~R.~Ellis, U.~W.~Heinz and H.~Kowalski,
Phys.\ Lett.\ {\bf B233}, 223 (1989).

\bibitem{Taylor}
T.~C.~Brooks {\it et al.}  [MiniMax Collaboration],
Phys.\ Rev.\ D {\bf 55}, 5667 (1997); hep-ph/9609375; Phys.\ Rev.\ D
{\bf 61}, 032003 (2000); hep-ex/9906026.

\bibitem{Pruneau} S.~Gavin, C.~Pruneau and S.~A.~Voloshin, in progress. 

\bibitem{Whitmore}
J.~Whitmore,
Phys.\ Rep.\  {\bf 27}, 187 (1976).

\bibitem{Boggild}
H.~Boggild and T.~Ferbel,
Ann.\ Rev.\ Nucl.\ Part.\ Sci.\  {\bf 24}, 451 (l974).

\bibitem{Mrowczynski} S.~Mrowczynski, nucl-th/0112007.

\bibitem{HIJING}
X.~N.~Wang and M.~Gyulassy,
Phys.\ Rev.\ D {\bf 44}, 3501 (1991);
S.~E.~Vance and M.~Gyulassy,
Phys.\ Rev.\ Lett.\  {\bf 83}, 1735 (1999); nucl-th/9901009; 
S.~E.~Vance, M.~Gyulassy and X.~N.~Wang, 
Phys.\ Lett.\ {\bf B443}, 45 (1998); nucl-th/9806008. 

\bibitem{hamlet} S.~Gavin, in progress.



\end{thebibliography}
\end{document}